\documentstyle[12pt]{article}
\newcommand{\sptwo}{1.4}

\newcommand{\doublespace}{\edef\baselinestretch{\sptwo}\Large\normalsize}

\begin{document}
\doublespace
\begin{center}
{\bf Separable Structure of Many-Body Ground-State Wave Function }\\
\renewcommand\thefootnote{\fnsymbol{footnote}}
{Yeong E. Kim \footnote{ e-mail:yekim$@$physics.purdue.edu} and
Alexander L. Zubarev\footnote{ e-mail: zubareva$@$physics.purdue.edu}}\\
Department of Physics, Purdue University\\
West Lafayette, Indiana  47907\\
{\bf Abstract}
\end{center}
\begin{quote}
We have investigated a general structure of the ground-state wave function for
 the Schr\"odinger
equation for $N$ identical interacting particles (bosons or fermions)
confined in a harmonic anisotropic trap in the limit of large $N$.
It is shown that the ground-state wave function can be written in a separable 
form. As an example of its applications, this form is used to obtain the 
ground-state wave function
describing collective dynamics for $N$ trapped bosons interacting via contact
 forces.

\end{quote}

\noindent

\vspace{55 mm}

\pagebreak
The structure of the ground-state wave function for a many-body system is very
 important for theoretical understanding of recently observed Bose-Einstein 
condensation (BEC) [1] (the theoretical
aspects of the BEC are discussed in recent reviews [2])
 and other many body problems. 
The Ginzburg-Pitaevskii-Gross (GPG) equation [3] is most widely used to describe the experimental results for the BEC.
Recently, an alternative method of equivalent linear two-body (ELTB) equations 
for many body systems has been developed based on the variational principle [4,5]. In this paper, we
consider $N$ identical particles (bosons or fermions) confined in a harmonic
 anisotropic trap. We show that in the case of large $N$ the ground-state wave 
function can be written in separable form as
$$
\Psi(\vec{r}_1, \vec{r}_2,...\vec{r}_N) = \phi(x, y, z) \cdot \chi (\Omega, 
\sigma),
\eqno{(1)}
$$
 
\noindent
where
$$
\everymath={\displaystyle}
x = \sqrt{\sum_{i=1}^N~x_i^2}, ~~ y = \sqrt{\sum_{i=1}^N~y_i^2}, ~~ z =
\sqrt{\sum_{i=1}^N~z_i^2},
\eqno{(2)}
$$

\noindent
$\Omega$ is a set of (3N - 3) angular variables, and $\sigma$ is a set of spin 
variables.

We start from a generalization of the hyperspherical expansion of the wave function for the Hamiltonian
$$
H=-\frac{\hbar^2}{2m}  \sum_{i=1}^{N} \Delta_{i}+\frac{1}{2}
m \sum_{i=1}^{N}(\omega_x^2x_i^2+\omega_y^2y_i^2+\omega_z^2z_i^2)
+\sum_{i<j}V_{int}({\bf r}_i-{\bf r}_j)
\eqno{(3)}
$$
in the form [4,6]
$$
\Psi({\bf r}_1, ...{\bf
r}_N)=\sum_{[K]}
\Phi_{[K]}(x,y,z)Y_{[K]}(\Omega^N_x, \Omega^N_y, \Omega^N_z, \sigma),
\eqno{(4)}
$$
where $Y_{[K]}(\Omega^N_x, \Omega^N_y, \Omega^N_z, \sigma)=Y_{K_x,K_y,K_z}^{\nu_x,\nu_y,\nu_z}(\Omega^N_x,\Omega^N_y,
\Omega^N_z, 
\sigma)$
 is the combination of the hyperspherical harmonics, $
Y^{\nu_x}_{K_x}(\Omega^N_x), Y^{\nu_y}_{K_y}(\Omega^N_y),$ and $Y^{\nu_z}_{K_z}
(\Omega^N_z)$,
with functions of spin variables $\sigma$,
which is symmetric or antisymmetric with respect to permutations of particles 
for bosons or fermions respectively. $[K]$ represents a set of numbers 
$[K_x,\nu_x,K_y,\nu_y,K_z,\nu_z]$. 

The hyperspherical harmonics $Y^{\nu_x} _{K_x}(\Omega^N_x)$, $
Y^{\nu_y}_{K_y}(\Omega^N_y)$, and $Y^{\nu_z}_{K_z}(\Omega^N_z)$
are eigenfunctions of the hyperspherical
angular parts of the Laplace operators $\sum_{i=1}^{N} \frac{\partial^2}
{\partial x_i^2}$, $\sum_{i=1}^{N} \frac{\partial^2}{\partial
y_i^2}$, and $\sum_{i=1}^{N} \frac{\partial^2}{\partial z_i^2}$,
respectively.

The Laplace operators are defined by
$$
\sum_{i=1}^{N} \frac{\partial^2}{\partial x_i^2}=\frac{1}{x^{N-1}}
\frac{\partial}{\partial x}(x^{N-1}\frac{\partial}{\partial x})+
\frac{1}{x^2} \Delta_{\Omega_x^N},
$$
$$
\sum_{i=1}^{N} \frac{\partial^2}{\partial y_i^2}=\frac{1}{y^{N-1}}
\frac{\partial}{\partial y}(y^{N-1}\frac{\partial}{\partial y})+
\frac{1}{y^2} \Delta_{\Omega_y^N},
\eqno{(5)}
$$
and
$$
\sum_{i=1}^{N} \frac{\partial^2}{\partial z_i^2}=\frac{1}{z^{N-1}}
\frac{\partial}{\partial z}(z^{N-1}\frac{\partial}{\partial z})+
\frac{1}{z^2} \Delta_{\Omega_z^N}.
$$

The hyperspherical angles $\theta_1^{x}, \theta_2^{x},...\theta_{N-1}^{x},
\theta_1^{y}, \theta_2^{y},...\theta_{N-1}^{y},
\theta_1^{z}, \theta_2^{z},...\theta_{N-1}^{z}$
can be chosen in such a way that the hyperspherical
angular parts of the Laplace operators $\Delta_{\Omega_t^N}$
satisfy the recursion relation [7]
$$
\Delta_{\Omega_u^N}=\frac{1}{\sin^{N-2} \theta^u_{N-1}}
\frac{\partial}{\partial\theta^u_N}(\sin^{N-2}\theta^u_{N-1}
\frac{\partial}{\partial\theta^u_{N-1}})+
\frac{1}{\sin^2\theta^u_{N-1}} \Delta_{\Omega_u^{N-1}}
\eqno{(6)}
$$
with $u=x,y,$ or $z$.

Functions $\Phi_{[K]}(x,y,z)$ satisfy equations

$$
\sum_{[K^{\prime}]}
h_{[K],[K^{\prime}]}
\Phi_{[K^{\prime}]}(x,y,z)=E\Phi_{[K]}(x,y,z),
\eqno{(7)}
$$
where
$$
\everymath={\displaystyle}
\begin{array}{rcl}
&~& h_{[K] [K^{\prime}]}=
\delta_{K_xK^{\prime}_x}\delta_{K_yK^{\prime}_y}\delta_{K_zK^{\prime}_z}
\delta_{\nu_x\nu^{\prime}_x}\delta_{\nu_y\nu^{\prime}_y}\delta_{\nu_z
\nu^{\prime}_z}[-\frac{\hbar^2}{2m}(\frac{\partial^2}{\partial x^2}
+\frac{\partial^2}{\partial y^2}+\frac{\partial^2}{\partial z^2})\\
&~& \\
&~& +\frac{m}{2}(\omega_x^2x^2+\omega_y^2y^2+\omega_z^2z^2)
+\frac{\hbar^2}{2m}(\frac{(N-1+2K_x)(N-3+2K_x)}{4x^2}\\
&~& \\
&~&+\frac{(N-1+2K_y)(N-3+2K_y)}{4y^2}
+\frac{(N-1+2K_z)(N-3+2K_z)}{4z^2}]\\
&~& \\
&~& + V_{[K][K^{\prime}]}(x,y,z),
\end{array}
\eqno{(8)}
$$
 with
$$
V_{[K] [K^{\prime}]
}(x,y,z)=
<K_x,\nu_x,K_y,\nu_y,K_z,\nu_z\mid\sum_{i<j}V_{int}({\bf r}_i-{\bf r}_j)
\mid K^{\prime}_x,\nu^{\prime}_x,K^{\prime}_y,\nu^{\prime}_y,K^{\prime}_z,\nu^{\prime}_z>.
\eqno{(9)}
$$

We write $\Phi_{[K]}(x,y,z)$ in the form of a Laplace integral
$$
\Phi_{[K]}(x,y,z)=\int f_{[K]}(\alpha_x,\alpha_y,\alpha_z)
\phi_x(x,\alpha_x)\phi_y(y,\alpha_y)\phi_z(z,\alpha_z)
d\alpha_x d\alpha_y d\alpha_z,
\eqno{(10)}
$$
where
$$
\phi_t(t, \alpha_t)=\sqrt{\frac{2}{\Gamma(N/2)}}(\frac{m \tilde{\omega}}
{\alpha_t^2 \hbar})^{N/4} \exp[-m \tilde{\omega}(\frac{t}{\alpha_t})^2
/(2\hbar)]t^{(N-1)/2},
\eqno{(11)}
$$
and $\tilde{\omega}=(\omega_x\omega_y\omega_z)^{1/3}$.

The Hill-Wheeler type equations [8,9] are obtained by requiring that energy of the system is stationary with respect to the functions $f_{[K]}(\alpha_x,\alpha_y,\alpha_z)$

$$
\everymath={\displaystyle}
\begin{array}{rcl}
&~& \sum_{[K]} \int d\alpha_x d\alpha_y d\alpha_z f_{[K]}(\alpha_x,\alpha_y,\alpha_z)[H_{[K][K^{\prime}]}(\alpha_x\alpha_y\alpha_z,\alpha^{\prime}_x
\alpha^{\prime}_y\alpha^{\prime}_z)\\
&~& \\
&~& -
\delta_{[K][K^{\prime}]}S(\alpha_x\alpha_y\alpha_z,\alpha^{\prime}_x
\alpha^{\prime}_y\alpha^{\prime}_z)E]=0,
\end{array}
\eqno{(12)}
$$

where
$$
\everymath={\displaystyle}
\begin{array}{rcl}
&~& H_{[K][K^{\prime}]}(\alpha_x \alpha_y \alpha_z,\alpha^{\prime}_x
 \alpha^{\prime}_y \alpha^{\prime}_z)= 
<\phi_x(x,\alpha_x)\phi_y(y,\alpha_y)\phi_z(z,\alpha_z)
Y_{[K]}\\
&~& \\
&~& \times \mid H \mid \phi_x(x,\alpha^{\prime}_x)\phi_y(y,\alpha^{\prime}_y)
\phi_z(z,\alpha^{\prime}_z)Y_{[K^{\prime}]}>,
\end{array}
\eqno{(13)}
$$

and
$$
S(\alpha_x\alpha_y\alpha_z,\alpha^{\prime}_x
\alpha^{\prime}_y \alpha^{\prime}_z)=
<\phi_x(x,\alpha_x)\phi_y(y,\alpha_y)\phi_z(z,\alpha_z)\mid
\phi_x(x,\alpha^{\prime}_x)\phi_y(y,\alpha^{\prime}_y)
\phi_z(z,\alpha^{\prime}_z)>.
\eqno{(14)}
$$

In order to solve the Hill-Wheeler type equations (12),
we assume that the integral in Eq. (10) can be replaced by sum
$$
\Phi_{[K]}(x,y,z)=\sum_{i,j,k=1}^{\infty} c^{[K]}_{ijk}
\phi_x(x,\alpha^i_x)\phi_y(y,\alpha^j_y)\phi_z(z,\alpha^k_z),
\eqno{(15)}
$$

where $c^{[K]}_{ijk}$
are solutions of the following equations
$$
 \sum_{\stackrel{i^{\prime},j^{\prime},k^{\prime}}{[K^{\prime}]}}
[H_{[K][K^{\prime}]}(\alpha^{i}_x\alpha^{j}_y\alpha^{k}_z,\alpha^{i^{\prime}}_x
\alpha^{j^{\prime}}_y\alpha^{k^{\prime}}_z)
-\delta_{[K][K^{\prime}]}S(\alpha^{i}_x\alpha^{j}_y\alpha^{k}_z,\alpha^{i^{\prime}}_x\alpha^{j^{\prime}}_y\alpha^{k^{\prime}}_z)E]
c^{[K^{\prime}]}_{i^{\prime}j^{\prime}k^{\prime}}=0.
\eqno{(16)}
$$

For the case of large $N$,
the overlap, Eq. (14), $$S(\alpha^{i}_x\alpha^{j}_y\alpha^{k}_z,\alpha^{i^{\prime}}_x\alpha^{j^{\prime}}_y\alpha^{k^{\prime}}_z)=
[\frac{8 \alpha^{i}_x \alpha^{i^{\prime}}_x
\alpha^{j}_y \alpha^{j^{\prime}}_y
\alpha^{k}_z \alpha^{k^{\prime}}_z}{((\alpha^{i}_x)^2+(\alpha^{i^{\prime}}_x)^2)
((\alpha^{j}_y)^2+(\alpha^{j^{\prime}}_y)^2)
((\alpha^{k}_z)^2+(\alpha^{k^{\prime}}_z)^2)}]^{N/2} 
\eqno{(17)}
$$
reduces to the Kronecker deltas 
$$
S(\alpha^{i}_x\alpha^{j}_y\alpha^{k}_z,\alpha^{i^{\prime}}_x
\alpha^{j^{\prime}}_y\alpha^{k^{\prime}}_z)= \delta_{ii^{\prime}}\delta_{jj^{\prime}}\delta_{kk^{\prime}}
\eqno{(18)}
$$

 Since the ratio
$$
H_{[K][K^{\prime}]}(\alpha^{i}_x\alpha^{j}_y\alpha^{k}_z,\alpha^{i^{\prime}}_x
\alpha^{j^{\prime}}_y\alpha^{k^{\prime}}_z)/
S(\alpha^{i}_x\alpha^{j}_y\alpha^{k}_z,\alpha^{i^{\prime}}_x\alpha^{j^{\prime}}_y\alpha^{k^{\prime}}_z)$$
is a much more slowly varying function of $\alpha$ compared to
$S(\alpha^{i}_x\alpha^{j}_y\alpha^{k}_z,\alpha^{i^{\prime}}_x\alpha^{j^{\prime}}
y\alpha^{k^{\prime}}_z)$ in almost all cases [10], we have for the case of large $N$
$$
H_{[K][K^{\prime}]}(\alpha^{i}_x\alpha^{j}_y\alpha^{k}_z,\alpha^{i^{\prime}}_x
\alpha^{j^{\prime}}_y\alpha^{k^{\prime}}_z)=\tilde{H}_{[K][K^{\prime}]}(
\tilde{\alpha_x},\tilde{\alpha_y},\tilde{\alpha_z}) 
 \delta_{ii^{\prime}}\delta_{jj^{\prime}}\delta_{kk^{\prime}},
\eqno{(19)}
$$
(see Appendix for the case of $N$ identical particles interacting via contact 
repulsive force).

Substitution of Eq. (19) into Eq. (16) gives
$$
\Phi_{[K]}(x,y,z)=\tilde{c}_{[K]}\phi_x(x,\tilde{\alpha_x})
\phi_y(y,\tilde{\alpha_y})\phi_z(z,\tilde{\alpha_z}),
\eqno{(20)}
$$
where $\tilde{c}_{[K]}$ are solutions of the following equations
$$
\sum_{[K^{\prime}]}[\tilde{H}_{[K][K^{\prime}]}(\tilde{\alpha_x},
\tilde{\alpha_y},\tilde{\alpha_z})- \delta_{[K][K^{\prime}]}E]
\tilde{c}_{[K^{\prime}]},
\eqno{(21)}
$$
and parameters $ \tilde{\alpha_x}, \tilde{\alpha_y}$, and $\tilde{\alpha_z}$
are solutions of
$$
\frac{\partial E}{\partial \tilde{\alpha_x}}=
\frac{\partial E}{\partial \tilde{\alpha_y}}=
\frac{\partial E}{\partial \tilde{\alpha_z}}=0.
$$
Substitution of Eq. (20) into Eq. (4) yields 
$\Psi(\vec{r}_1, \vec{r}_2,...\vec{r}_N)$ given by Eq. (1) with

$$
\phi(x,y,z)=\phi_x(x,\tilde{\alpha_x})
\phi_y(y,\tilde{\alpha_y})\phi_z(z,\tilde{\alpha_z}),
$$
and
$$
\chi(\Omega,\sigma)=\sum_{[K]}\tilde{c}^{[K]}Y_{[K]}(\Omega^N_x,\Omega^N_y,
\Omega^N_z,\sigma).
$$

We now consider $N$ identical particles confined in an anisotropic harmonic trap and interacting via contact force
$$
V_{int}(\vec{r}_i-\vec{r}_j)=\frac{4 \pi \hbar^2a}{m}\delta(\vec{r}_i-\vec{r}_j)
,
\eqno{(23)}
$$
with positive scattering length $a>0$.
Using factorization (1) we have
$$
\begin{array}{rcl}
&~& [-\frac{\hbar^2}{2m}(\frac{\partial^2}{\partial x^2}+\frac{\partial
^2}
{\partial y^2}+\frac{\partial^2}{\partial z^2})+\frac{m}{2}(\omega_x^2x^2
+\omega_y^2y^2+\omega_z^2z^2)-\frac{\hbar^2}{2m}(
\frac{c_x}{x^2}+\frac{c_y}{y^2}+\frac{c_z}{z^2})\\
&~& \\
&~& +\frac{\hbar^2}{2m}\frac{(N-1)(N-3)}{4}(
\frac{1}{x^2}+\frac{1}{y^2}+\frac{1}{z^2})
+\frac{c}{xyz}]\phi(x,y,z)=E \phi(x,y,z),
\end{array}
\eqno{(24)}
$$
where $c_t=<\chi\mid\Delta_{\Omega_t^N}\mid\chi>/<\chi\mid\chi>$ with $t=(x,y,z)$
and
$$c=\frac{a\hbar^2N(N-1)}{\sqrt{2\pi}m}(\frac{\Gamma(N/2)}{\Gamma((N-1)/2)})^{3}
\tilde{c}.$$
 In the large $N$ limit, parameters $c_x, c_y, c_z,$ and  $\tilde{c}
$ are expected to be slowly varying functions of $N$.
For $N$ identical bosonic atoms with large $N,$ an
essentially exact expression for the ground state energy can be obtained
by neglecting the kinetic energy term in the
GPG equation [2,3] (this is called ``Thomas-Fermi
approximation" [11]).
From comparison of the ground state solution of Eq. (24) with the Thomas-Fermi
approximation [11], we can fix unknown parameters and find the ground-state
solution of Eq. (24) as
$$
\phi(x,y,z)=\psi_x(x)\psi_y(y)\psi_z(z),
\eqno{(25)}
$$
$$
E=\frac{5 N \hbar \tilde{\omega}}{4} \tilde{n}^{2/5},
\eqno{(26)}
$$
with
$$
\everymath={\displaystyle}
\begin{array}{rcl}
\psi_x(x)=Ax^{(N-1)/2}\exp[-m\tilde{\omega}(x/\alpha)^2/(2 \hbar)],\\
\psi_y(y)=Ay^{(N-1)/2}\exp[-m\tilde{\omega}(y/\beta)^2/(2 \hbar)],\\
\psi_z(z)=Az^{(N-1)/2}\exp[-m\tilde{\omega}(z/\gamma)^2/(2 \hbar)],
\end{array}
\eqno{(27)}
$$
where $A=\sqrt{2/\Gamma(N/2)}(m\tilde{\omega}/(\alpha^2 \hbar))^{N/4}$,
$\alpha=\tilde{n}^{1/5}\tilde{\omega}/\omega_x$,
$\beta=\tilde{n}^{1/5}\tilde{\omega}/\omega_y$,
$\gamma=\tilde{n}^{1/5}\tilde{\omega}/\omega_z$,
$\tilde{\omega}=(\omega_x\omega_y\omega_z)^{1/3}, \tilde{n}=n\tilde{c},
n=2\sqrt{\tilde{\omega}m/(2\pi\hbar)}Na$
and
$$
\tilde{c}=(\frac{4}{7})^{5/2}\frac{15}{8}\sqrt{\pi}\approx 0.82.
\eqno{(28)}
$$
Eqs. (25-28) give the exact ground-state solution of Eq. (24)
for large $N$. Thus we have found an analytical solution for the 
ground-state wave function
describing collective dynamics in variables (x,y,z) in the large $N$ limit.

We note that the slope of the Thomas-Fermi wave function becomes infinity at
 the surface, leading to logarithmic singularity in the kinetic energy. Hence 
it is necessary to modify the Thomas-Fermi wave function near the surface 
[12-14].
In contrast, we do not have such problems
for our solution, Eq. (25-28).

It is also interesting to compare our results with the ELTB method [4,5].
For this situation (contact force, Eq. (23) and large N limit), the ELTB method
corresponds to $\tilde{c}^{2/5}=1$. It shows that the ELTB method is a very good 
approximation with relative error of about $8 \%$ for parameter $\tilde{c}^{2/5}$.

In summary, we have investigated the general structure of the 
ground-state solution of 
the Schr\"odinger
equation for $N$ identical interacting particles (bosons or fermions)
confined in a harmonic anisotropic trap in the large $N$ limit.
 The main results and conclusions are as follows

(i) It has been shown that in the case of large $N$ the ground-state wave 
function
can be written in separable form, Eq. (1).

(ii) Using this form, we have found an analytical solution for the
ground-state wave function, 
Eqs. (25-29), describing collective dynamics in collective variables
(x,y,z)
for $N$ trapped bosons interacting via contact repulsive forces. 

(iii) Our results can be used for checking
various approximations (both existing and future) made for the Schr\"odinger
 equation describing $N$ identical interacting particles (bosons or fermions)
confined in a harmonic anisotropic trap.

\pagebreak

\begin{center}
{\bf Appendix}
\end{center}
To prove Eq. (19) we consider the contact potential case
$$
V_{int}({\bf r}_i-{\bf r}_j,\sigma)=\delta({\bf r}_i-{\bf r}_j)\eta(\sigma),
\eqno{(A.1)}
$$
where $\eta$ depends on spin variables.
Using Eq. (A.1) we can rewrite Eq. (9) as
$$
V_{[K][K^{\prime}]}(x,y,z)=\gamma_{[K][K^{\prime}]}\frac{N(N-1)}{xyz},
\eqno{(A.2)}
$$
where $\gamma_{[K][K^{\prime}]}$ does not depend on $x,y,z$.

Substitution of Eq. (A.2) into Eq. (13) gives
$$
\everymath={\displaystyle}
\begin{array}{rcl}
&~& H_{[K][K^{\prime}]}(\alpha^{i}_x\alpha^{j}_y\alpha^{k}_z,\alpha^{i^{\prime}}_x
\alpha^{j^{\prime}}_y\alpha^{k^{\prime}}_z)/(\hbar \tilde{\omega} N)=(1/2)
S(\alpha^{i}_x\alpha^{j}_y\alpha^{k}_z,\alpha^{i^{\prime}}_x\alpha
^{j^{\prime}}_y\alpha^{k^{\prime}}_z)
\\
&~&\\
&~& \times[
\delta_{[K][K^{\prime}]}(\frac{1+(\alpha^{i}_x)^2
(\alpha^{i^{\prime}}_x)^2 \beta_x^2}{(\alpha^{i}_x)^2+(\alpha^{i^{\prime}}_x)^2}
+\frac{1+(\alpha^{j}_y)^2 (\alpha^{j^{\prime}}_y)^2 \beta_y^2}
{(\alpha^{j}_y)^2+(\alpha^{j^{\prime}}_y)^2}
+\frac{1+(\alpha^{k}_z)^2 (\alpha^{k^{\prime}}_z)^2 \beta_z^2}
{(\alpha^{k}_z)^2+(\alpha^{k^{\prime}}_z)^2})\\
&~& \\
&~& +\frac{\sqrt{((\alpha^{i}_x)^2+(\alpha^{i^{\prime}}_x)^2)
((\alpha^{j}_y)^2+(\alpha^{j^{\prime}}_y)^2)
((\alpha^{k}_z)^2+(\alpha^{k^{\prime}}_z)^2)
}}
{\alpha^{i}_x \alpha^{i^{\prime}}_x
\alpha^{j}_y \alpha^{j^{\prime}}_y
\alpha^{k}_z \alpha^{k^{\prime}}_z}
(\frac{\Gamma((N-1)/2)}{\Gamma(N/2)})^3\\
&~& \\
&~& \times(\frac{m\tilde{\omega}}{\hbar})^{3/2}\frac{N-1}{\sqrt{8}}\gamma_{[K][K^{\prime}]}],
\end{array}
\eqno{(A.3)}
$$
where $\beta_t=\omega_t/\tilde{\omega}$ for $t=x,y$, or $z$.

For large $N$,  $S(\alpha^{i}_x\alpha^{j}_y\alpha^{k}_z,\alpha^{i^{\prime}}_x\alpha
^{j^{\prime}}_y\alpha^{k^{\prime}}_z)$, Eq. (14), reduces to the Kronecker 
deltas
$\delta_{ii^{\prime}}
\delta_{jj^{\prime}}
\delta_{kk^{\prime}}$, and hence from Eq. (A.3) we obtain Eq. (19).

\pagebreak

\begin{center}
{\bf References}
\end{center}

\vspace{8pt}
\noindent
1. http://amo.phy.gasou.edu/bec.html, and references therein.

\noindent
2. K. Burnett, M. Edwards, and C.W. Clark, Phys. Today, {\bf 52}, 37 (1999);
 F. Dalfovo, S. Giorgini, L. Pitaevskii, and S. Stringari, Rev. Mod. Phys.
 {\bf 71}, 463 (1999). 

\noindent
3. L. Ginzburg and L.P. Pitaevskii, Zh. Eksp. Teor. Fiz. {\bf 34}, 1240 (1958)
[Sov. Phys. JETP, {\bf 858} (1958)];
E.P. Gross, J. Math. Phys. {\bf 4}, 195 (1963).

\noindent
4. A.L. Zubarev and Y.E. Kim, Phys. Lett. A{\bf 263}, 33 (1999).

\noindent
5. Y.E. Kim and A.L. Zubarev, J. Phys. B: At. Mol. Opt. Phys. {\bf 33}, 55 (2000).

\noindent
6. Yu. T. Grin', Yad. Fiz. {\bf 12}, 927 (1970) [Sov. J. Nucl. Phys. {\bf 12}, 345 (1971)].

\noindent
7. Yu. F. Smirnov and K. V. Shitikova, Fiz. Elem.
Chastits At. Yadra {\bf 8}, 847 (1977) [Sov. J. Part. Nucl. {\bf 8},
344 (1977)].

\noindent
8. D.L. Hill and J.A. Wheeler, Phys. Rev. {\bf 89}, 1102 (1953).

\noindent
9. J.J. Griffin and J.A. Wheeler, Phys. Rev. {\bf 108}, 311 (1957).

\noindent
10. D.M. Brink and A. Weiguny, Nucl. Phys. A{\bf 120}, 59 (1968).

\noindent
11. G. Baym and C. J. Pethick, Phys. Rev. Lett. {\bf 76}, 6
(1996).

\noindent
12. A.L. Fetter and D.L. Feder, Phys. Rev. A{\bf 58}, 3185 (1998).

\noindent
13. F. Dalfovo, L.P. Pitaevskii, and S. Stringari, Phys. Rev. A{\bf 54}, 4213 (1996).

\noindent
14. E. Lundh, C.J. Pethick, and H. Smith, Phys. Rev. A{\bf 55}, 2126 (1997).
\end{document}